\documentclass[a4paper]{jpconf}
\usepackage{graphicx}
\begin{document}
\title{Fluctuations of the azimuthal particle distribution\\ in NA49 at the CERN SPS}

\author{Tomasz Cetner and Katarzyna Grebieszkow\\for the NA49 Collaboration}

\address{Warsaw University of Technology, Faculty of Physics, Koszykowa 75, 00-662 Warszawa, Poland}

\ead{Tomasz.Cetner@cern.ch, kperl@if.pw.edu.pl}

\begin{abstract}
Event-by-event fluctuations and correlations in azimuthal angle are currently widely investigated in various experiments.
In this paper the $\Phi$ measure (earlier used in experiments to evaluate fluctuations in transverse momentum) is now applied to azimuthal angle 
$\phi$. Properties of this $\Phi_{\phi}$ function are investigated through fast generators and with complex models such as Pythia, Hijing, and UrQMD. Preliminary results of NA49 on $\Phi_{\phi}$ are also presented. 
The system size dependence (p+p, C+C, Si+Si and 6 centralities of Pb+Pb) at the highest SPS energy (158$A$ GeV) is shown, as well as the energy dependence (20$A$ - 158$A$ GeV) for the  7.2\% most central Pb+Pb interactions.

\end{abstract}

\vspace{0.5cm}
\noindent{\bf The NA49 Collaboration:}

\noindent
T.~Anticic$^{22}$, B.~Baatar$^{8}$, D.~Barna$^{4}$, J.~Bartke$^{6}$, 
H.~Beck$^{9}$, L.~Betev$^{10}$, H.~Bia{\l}\-kowska$^{19}$, C.~Blume$^{9}$, 
M.~Bogusz$^{21}$, B.~Boimska$^{19}$, J.~Book$^{9}$, M.~Botje$^{1}$,
P.~Bun\v{c}i\'{c}$^{10}$,
T.~Cetner$^{21}$, P.~Christakoglou$^{1}$,
P.~Chung$^{18}$, O.~Chv\'{a}la$^{14}$, J.G.~Cramer$^{15}$, V.~Eckardt$^{13}$,
Z.~Fodor$^{4}$, P.~Foka$^{7}$, V.~Friese$^{7}$,
M.~Ga\'zdzicki$^{9,11}$, K.~Grebieszkow$^{21}$, C.~H\"{o}hne$^{7}$,
K.~Kadija$^{22}$, A.~Karev$^{10}$, V.I.~Kolesnikov$^{8}$, M.~Kowalski$^{6}$, 
D.~Kresan$^{7}$,
A.~L\'{a}szl\'{o}$^{4}$, R.~Lacey$^{18}$, M.~van~Leeuwen$^{1}$,
M.~Ma\'{c}kowiak$^{21}$, M.~Makariev$^{17}$, A.I.~Malakhov$^{8}$,
M.~Mateev$^{16}$, G.L.~Melkumov$^{8}$, M.~Mitrovski$^{9}$, St.~Mr\'owczy\'nski$^{11}$, 
V.~Nicolic$^{22}$, G.~P\'{a}lla$^{4}$, A.D.~Panagiotou$^{2}$, W.~Peryt$^{21}$, 
J.~Pluta$^{21}$, D.~Prindle$^{15}$,
F.~P\"{u}hlhofer$^{12}$, R.~Renfordt$^{9}$, C.~Roland$^{5}$, G.~Roland$^{5}$,
M. Rybczy\'nski$^{11}$, A.~Rybicki$^{6}$, A.~Sandoval$^{7}$, 
N.~Schmitz$^{13}$, T.~Schuster$^{9}$, P.~Seyboth$^{13}$, F.~Sikl\'{e}r$^{4}$, 
E.~Skrzypczak$^{20}$, M.~S{\l}odkowski$^{21}$, G.~Stefanek$^{11}$, R.~Stock$^{9}$, 
H.~Str\"{o}bele$^{9}$, T.~Susa$^{22}$, M.~Szuba$^{21}$, 
M.~Utvi\'{c}$^{9}$, D.~Varga$^{3}$, M.~Vassiliou$^{2}$,
G.I.~Veres$^{4}$, G.~Vesztergombi$^{4}$, D.~Vrani\'{c}$^{7}$,
Z.~W{\l}odarczyk$^{11}$, A.~Wojtaszek-Szwarc$^{11}$

\vspace{0.5cm}
\noindent
$^{1}$ NIKHEF, Amsterdam, Netherlands. \\
$^{2}$ Department of Physics, University of Athens, Athens, Greece.\\
$^{3}$ E\"otv\"os Lor\'ant University, Budapest, Hungary \\
$^{4}$ KFKI Research Institute for Particle and Nuclear Physics, Budapest, Hungary.\\
$^{5}$ MIT, Cambridge, USA.\\
$^{6}$ H.~Niewodnicza\'nski Institute of Nuclear Physics, Polish Academy of Sciences, Cracow, Poland.\\
$^{7}$ Gesellschaft f\"{u}r Schwerionenforschung (GSI), Darmstadt, Germany.\\
$^{8}$ Joint Institute for Nuclear Research, Dubna, Russia.\\
$^{9}$ Fachbereich Physik der Universit\"{a}t, Frankfurt, Germany.\\
$^{10}$ CERN, Geneva, Switzerland.\\
$^{11}$ Institute of Physics, Jan Kochanowski University, Kielce, Poland.\\
$^{12}$ Fachbereich Physik der Universit\"{a}t, Marburg, Germany.\\
$^{13}$ Max-Planck-Institut f\"{u}r Physik, Munich, Germany.\\
$^{14}$ Inst. of Particle and Nuclear Physics, Charles Univ., Prague, Czech Republic.\\
$^{15}$ Nuclear Physics Laboratory, University of Washington, Seattle, WA, USA.\\
$^{16}$ Atomic Physics Department, Sofia Univ. St. Kliment Ohridski, Sofia, Bulgaria.\\
$^{17}$ Institute for Nuclear Research and Nuclear Energy, BAS, Sofia, Bulgaria.\\
$^{18}$ Department of Chemistry, Stony Brook Univ. (SUNYSB), Stony Brook, USA.\\
$^{19}$ Institute for Nuclear Studies, Warsaw, Poland.\\
$^{20}$ Institute for Experimental Physics, University of Warsaw, Warsaw, Poland.\\
$^{21}$ Faculty of Physics, Warsaw University of Technology, Warsaw, Poland.\\
$^{22}$ Rudjer Boskovic Institute, Zagreb, Croatia.\\

\section{$\Phi_{\phi}$ measure}

\subsection{Motivation}

The reasons to perform event-by-event azimuthal angle fluctuations were to investigate plasma instabilities \cite{plasma_inst}, search for the Critical Point and Onset of Deconfinement, and measure fluctuations of elliptic flow \cite{Mrow_flow_fluct, Mill_flow_fluct}.
In this work the $\Phi$ measure \cite{Phi} is chosen as it is a strongly intensive measure of fluctuations (does not depend on volume and on volume fluctuations). The $\Phi$ measure was already successfully used by NA49 to study average $p_T$ fluctuations \cite{phipt_syssize, phipt_energy} and charge fluctuations \cite{delta_q}. There are several other effects that influence the values of this measure, such as resonance decays, momentum conservation, flow, and quantum statistics. The $\Phi$ measure is not well suited to the analysis of these effects which represent a background for the present study.

\subsection{$\Phi_{\phi}$ definition}

Let $\phi$ be a particle's azimuthal production angle. One can define a single-particle variable $z_{\phi} \equiv \phi-\overline{\phi}$,
where $\overline{\phi}$ is an average over the single-particle inclusive azimuthal angle distribution.
Let us also define an event variable ${Z}_{\phi} \equiv \sum_{i=1}^{N}(\phi_i-\overline{\phi})$,
where the summation runs over particles in a given event.
Then the $\Phi_{\phi}$ measure is defined as:
\centerline{ ${\large \Phi_{\phi} \equiv \sqrt{\frac{\langle {Z}_{\phi}^{2} \rangle }{\langle N \rangle }}-\sqrt{\overline{z_{\phi}^{2}}} }$ ,}
where $\langle ... \rangle$ represents averaging over events.
When particles are emitted by a number ($N_s$) of identical sources, 
which are independent of each other and $P(N_s)$ is the distribution of this number, then $\Phi_{\phi}(N_s)$ is independent of $(N_s)$ (an intensive measure) and its distribution $P(N_s)$ (a strongly intensive measure).
Additionally, if particles are produced independently (no inter-particle correlations), $\Phi_{\phi}$ equals zero.

\section{Model studies}

General properties of the $\Phi_{\phi}$ measure were already studied using analytical calculations \cite{mrow_phiphi}. Fluctuations due to elliptic flow, quantum statistics and resonance decays in the hadron gas were evaluated. In this paper we present model studies that were conducted to investigate in detail the behavior of $\Phi_{\phi}$. Several toy models were employed to test separately the influence of various physics effects. Full event generators such as UrQMD \cite{urqmd}, Pythia \cite{pythia}, and Hijing \cite{hijing} were used as a direct reference to experimental results\footnote{see conference slides for Pythia and Hijing results}.

\subsection{Toy models of elliptic flow and momentum conservation}
\begin{figure}[h]
\hspace{1pc}%
\begin{minipage}{17pc}
\centerline{\includegraphics[width=17pc]{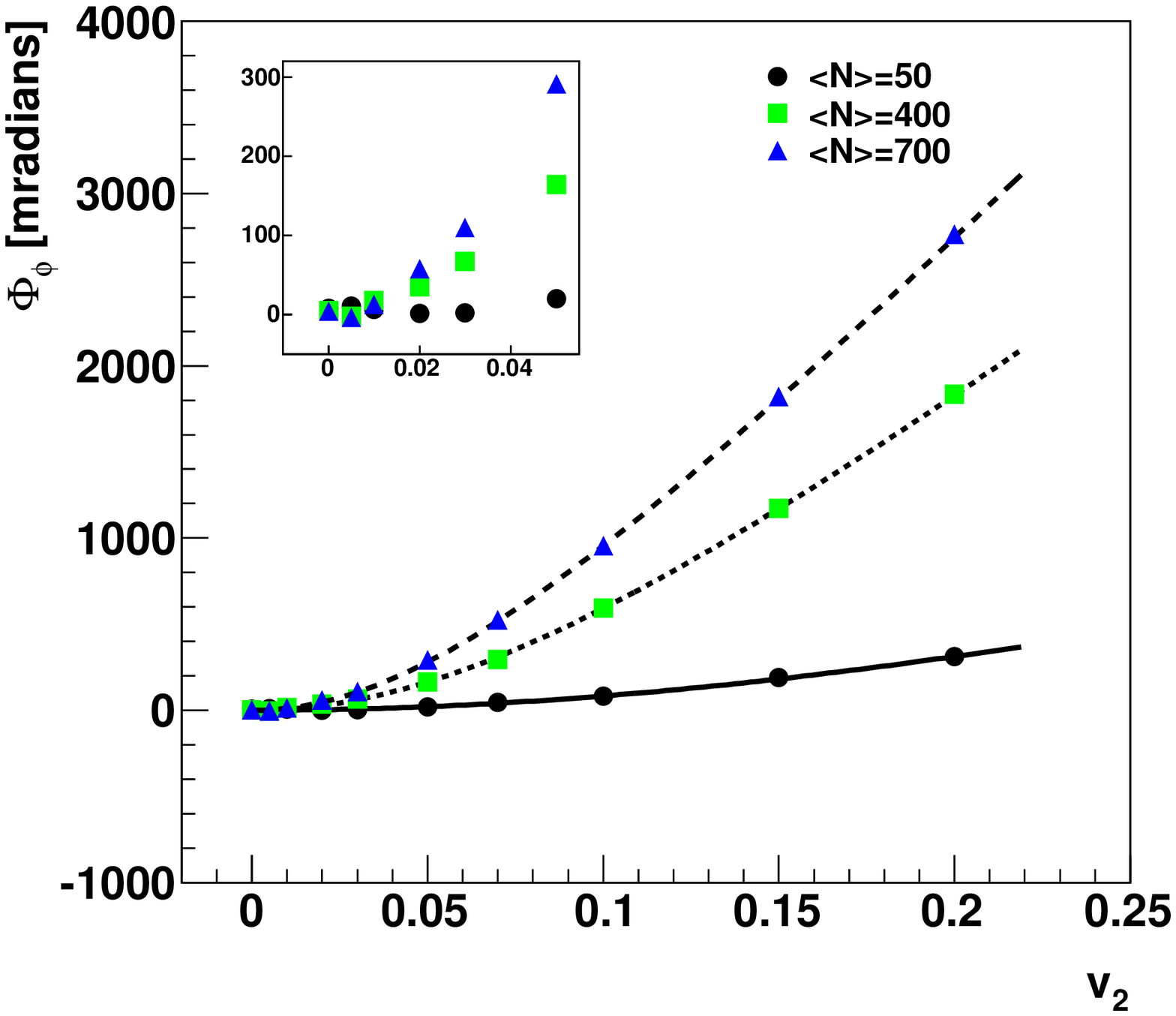}}
\vspace{-2pc}
\caption{\label{flow}\small $\Phi_\phi$ as a function of $v_2$ for elliptic flow simulation with constant $v_2$. Lines correspond to analytical formulas from \cite{mrow_phiphi}.}
\end{minipage}
\hspace{2pc}%
\begin{minipage}{17pc}
\centerline{\includegraphics[width=17pc]{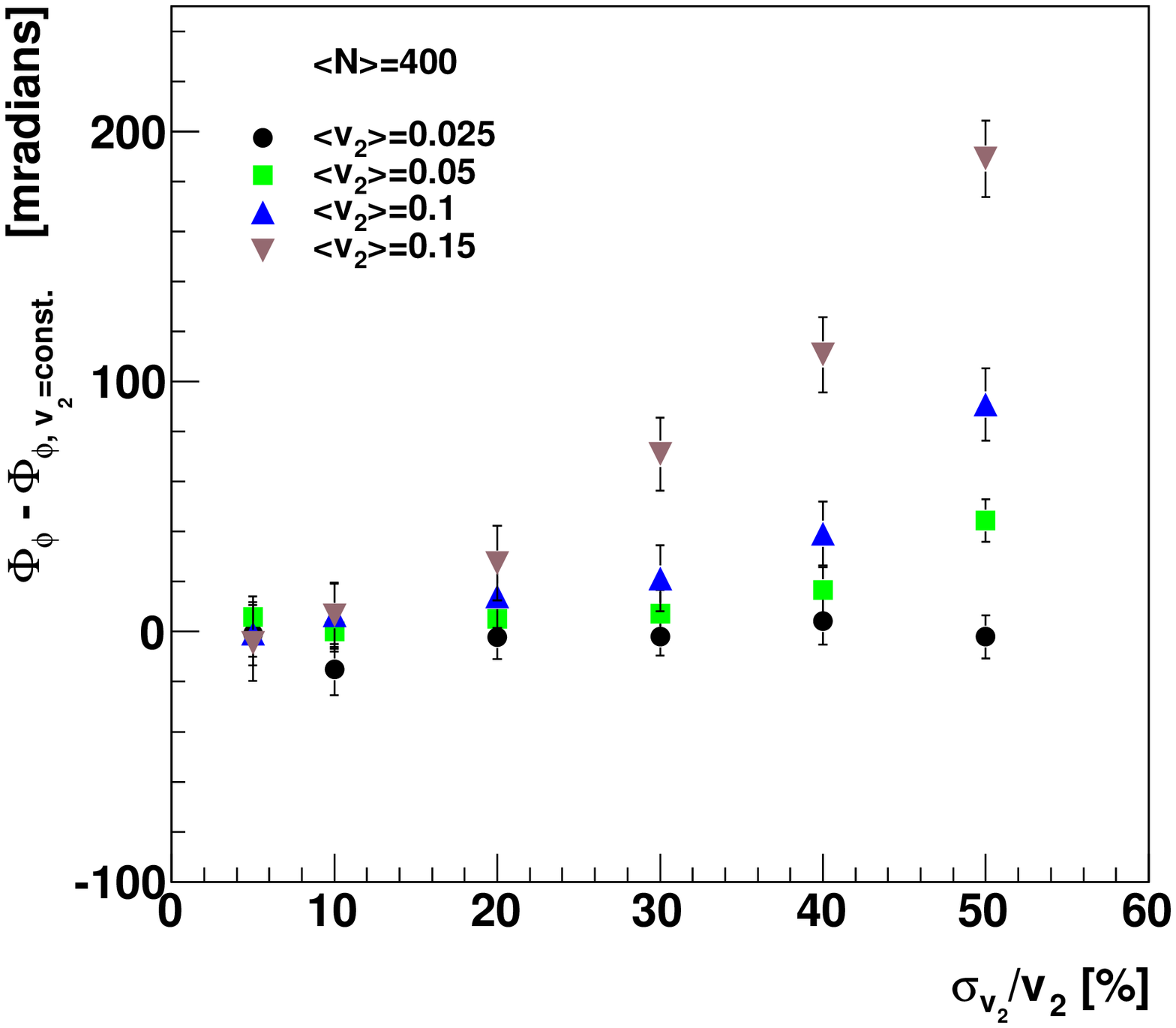}}
\vspace{-2pc}
\caption{\label{flow_fluct}\small Difference between $\Phi_\phi$ values for constant and fluctuating $v_2$ as a function of relative size of $v_2$ fluctuations.}
\end{minipage} 
\end{figure}

To simulate elliptic flow events were generated with azimuthal angle distribution of particles following $\rho(\phi) = 1+  2 v_2 cos(2(\phi - \phi_R))$. 
For each event the reaction plane angle $\phi_R$ was generated from a flat distribution in azimuthal angle, and the multiplicity N of particles in an event was taken from a Negative Binomial distribution 
with given $\langle N \rangle$ and dispersion $D_N \approx 0.5 \cdot \langle N \rangle$. The value of $v_2$ was a simulation parameter, constant for each simulation series (Fig. \ref{flow}) 
or varying from event to event according to a Gaussian distribution with $\sigma_{v_2}$  (Fig. \ref{flow_fluct}).

The elliptic flow effect results in positive $\Phi_{\phi}$ values which are increasing with increasing $v_2$. Moreover, event-by-event fluctuations in $v_2$ cause an additional increase of $\Phi_{\phi}$ values.

To simulate global momentum conservation (Fig. \ref{zachowanie}), 
each particle had its transverse momentum $p_T$ drawn from the distribution $P(p_T) \sim p_T e^{-p_T \over T}$, where $T=200$ MeV, and its azimuthal angle from a flat distribution.
Afterwards, for every particle $p_x$ and $p_y$ was modified: $p_x^{'} = p_x - \frac{\sum_{i=1}^{N}p_{x_i}}{N}$, and $p_y^{'} = p_y - \frac{\sum_{i=1}^{N}p_{y_i}}{N}$ to obey momentum conservation in the whole event ($N$ is the event multiplicity generated from a Negative Binomial distribution). 
Results show negative $\Phi_{\phi}$ values (anti-correlation) which weakly depend on multiplicity.

\subsection{UrQMD -- a full event generator}

\begin{figure}[h]
\hspace{1pc}%
\begin{minipage}{17pc}
\centerline{\includegraphics[width=17pc]{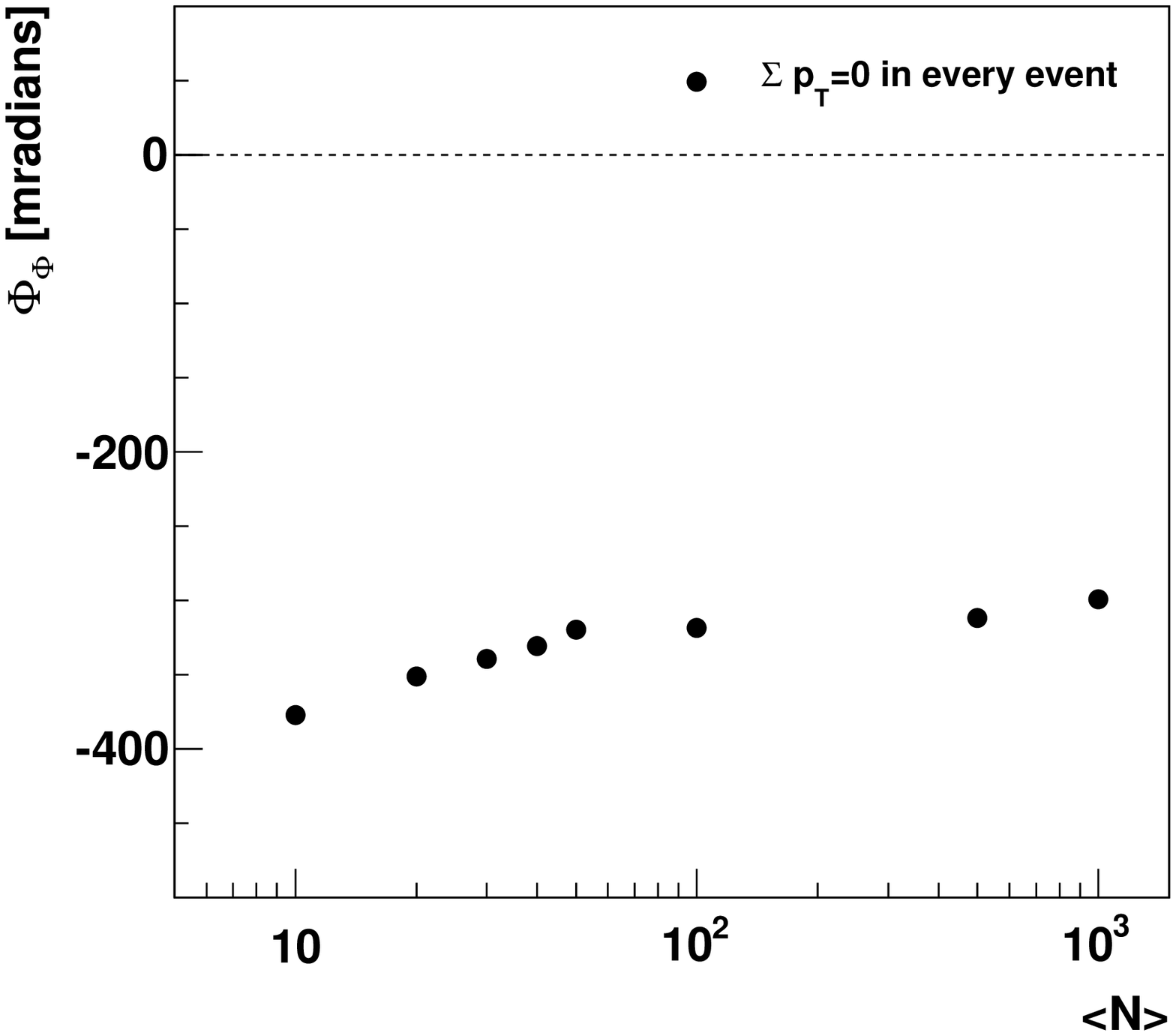}}
\vspace{-2pc}
\caption{\label{zachowanie}\small $\Phi_\phi$ as a function of mean multiplicity for simulation with momentum conservation.}
\end{minipage}\hspace{2pc}%
\begin{minipage}{17pc}
\centerline{\includegraphics[width=17pc]{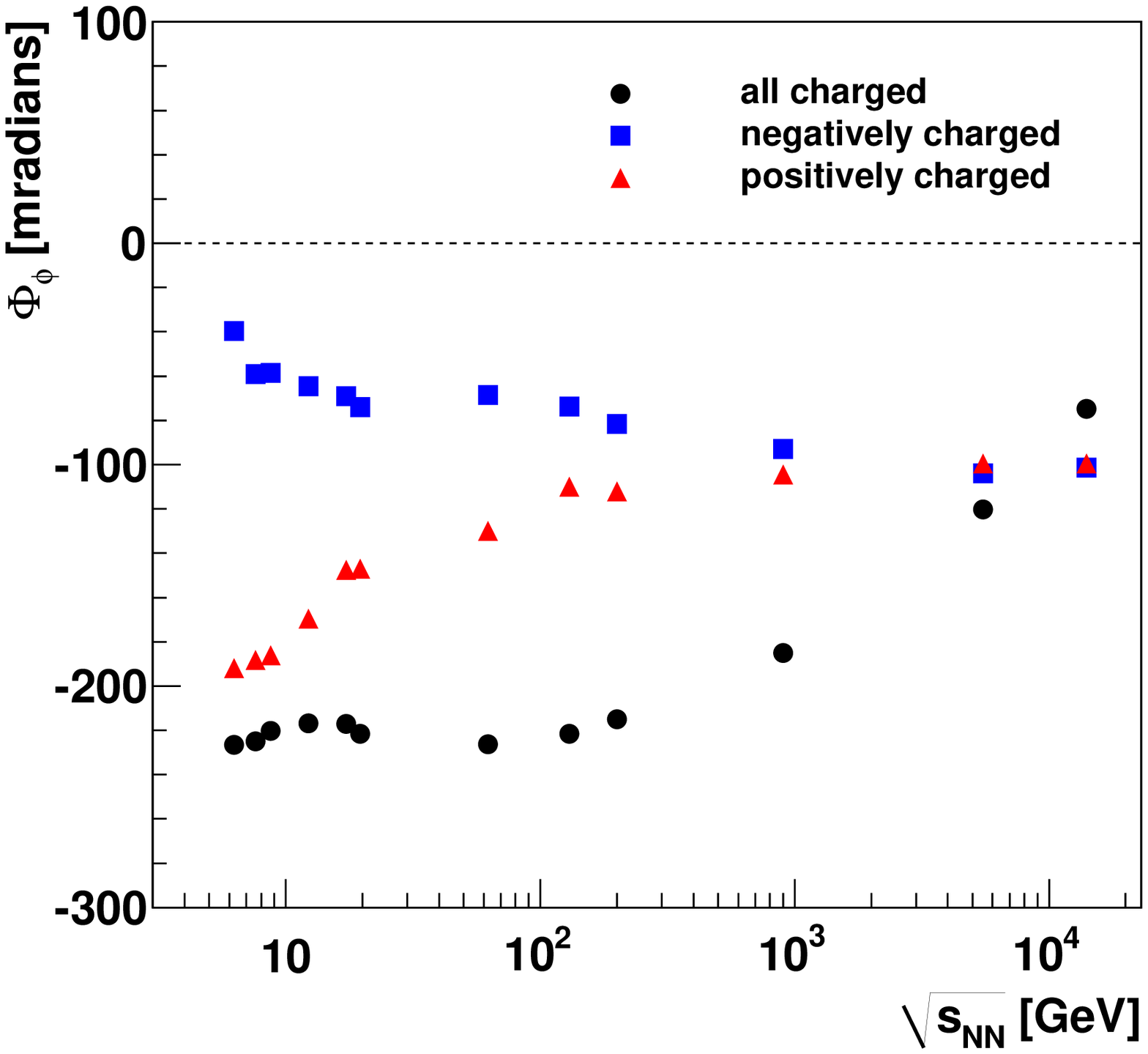}}
\vspace{-2pc}
\caption{\label{urqmd_pp}\small $\Phi_\phi$ as a function of energy for p+p collisions generated using the UrQMD 3.3 model.}
\end{minipage} 
\end{figure}

A simulation using UrQMD 3.3 was performed for p+p collisions in a wide energy range (SPS, RHIC, LHC) (Fig. \ref{urqmd_pp}). 
Default parameters of the generator were used, which for this version means simulation of hard processes (for energies $\sqrt{s}>10$ GeV), but no hydrodynamics stage.

For all energies and charge combinations $\Phi_\phi$ values are negative (domination of anti-correlations). Their magnitude increases for negatives and decreases for positives and all charged. At LHC energies the values become charge independent.

\section{NA49 experiment}

The NA49 experiment \cite{na49_nim} at the CERN SPS operated between 1994 and 2002. The registered data include p+p, C+C, Si+Si, and Pb+Pb interactions at a center of mass energy of 6.3~-~17.3~GeV per N+N pair. It is a fixed target hadron spectrometer (four TPCs, two of them, VTPC-1 and VTPC-2, inside a magnetic field) with precise determination of the collision centrality (Forward Calorimeter measuring energy of projectile spectators).

\subsection{Detector}
\begin{figure}[h]
\hspace{1pc}%
\begin{minipage}{17pc}
\centerline{\includegraphics[width=15pc]{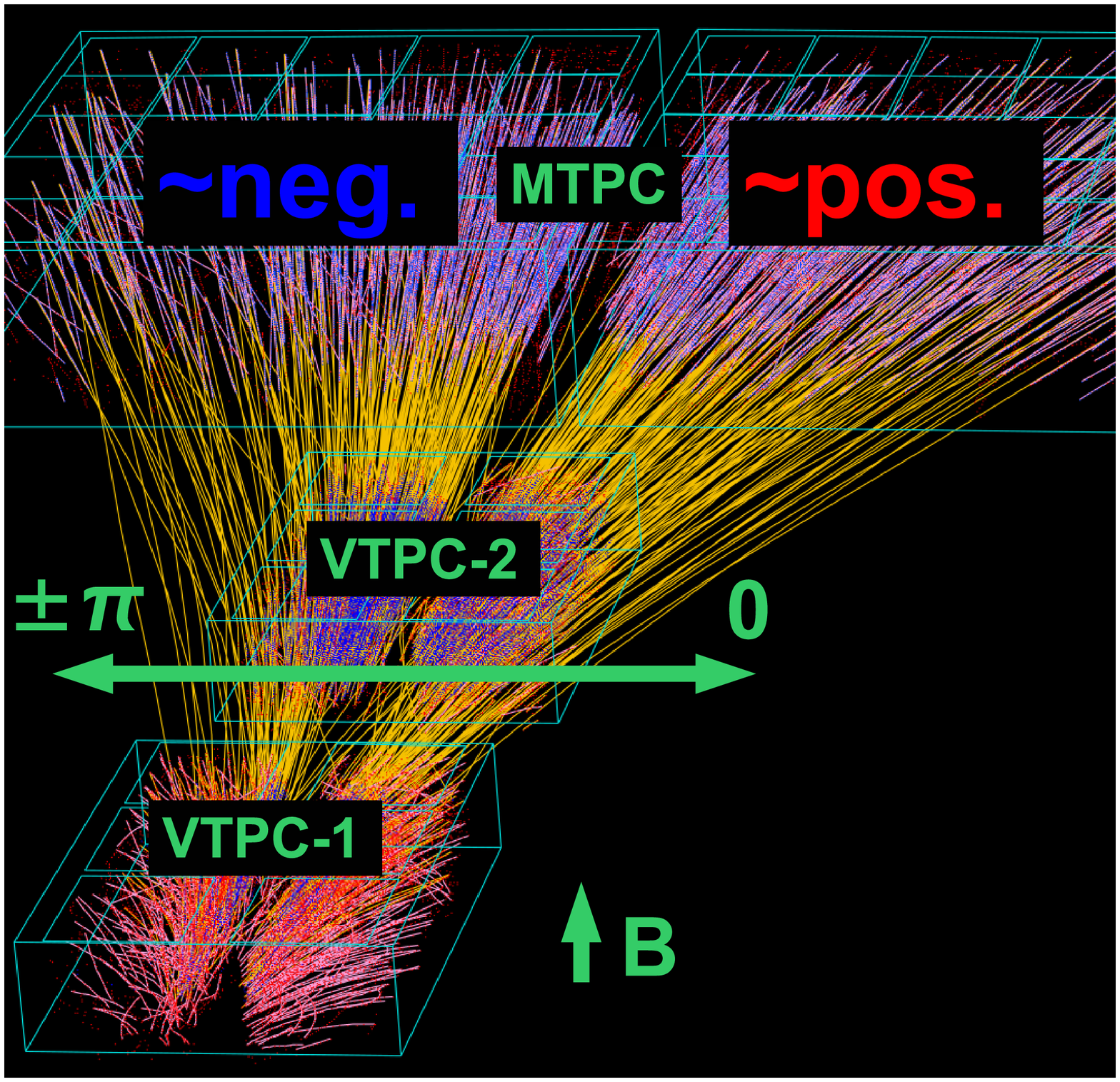}}
\vspace{-0.3pc}
\caption{\label{na49_az_acc}\small Tracks registered in NA49; four TPCs with marked regions where particles of given charge are dominant.}
\end{minipage}\hspace{2pc}%
\begin{minipage}{17pc}
\centerline{\includegraphics[width=15pc]{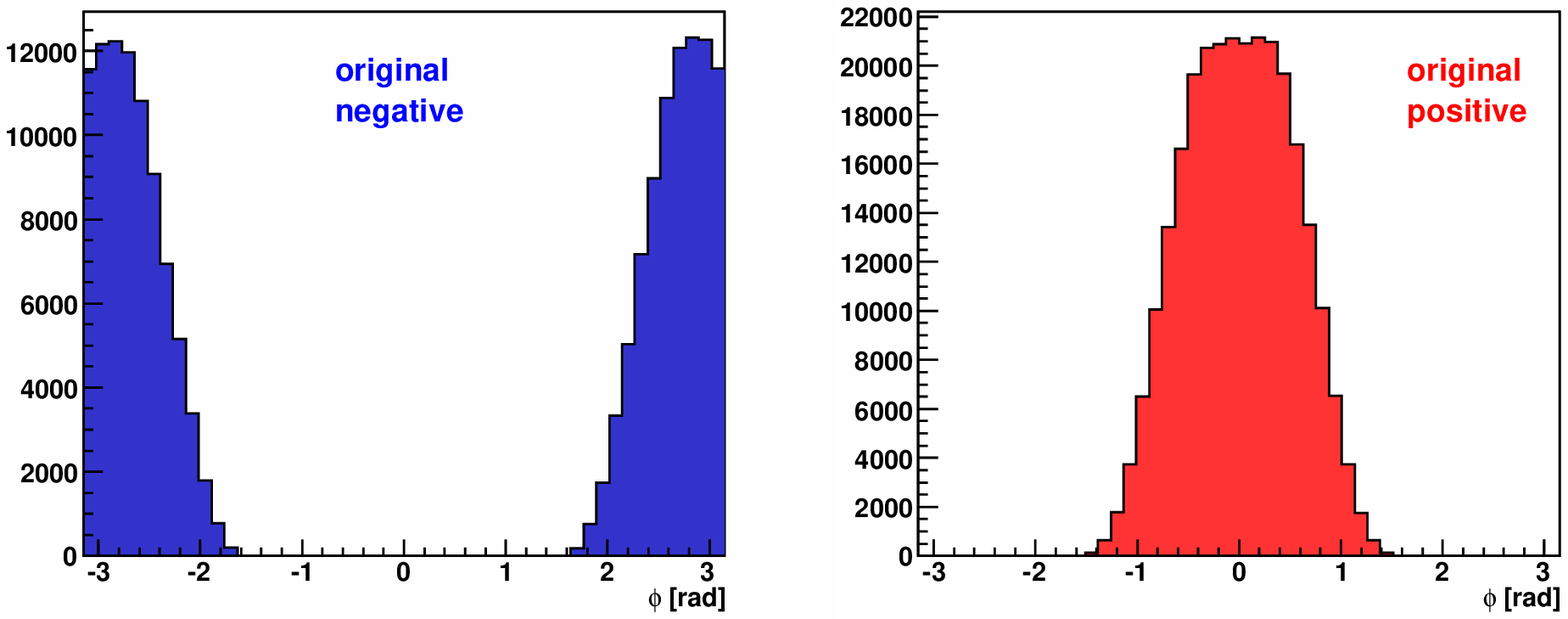}}
\centerline{\includegraphics[width=15pc]{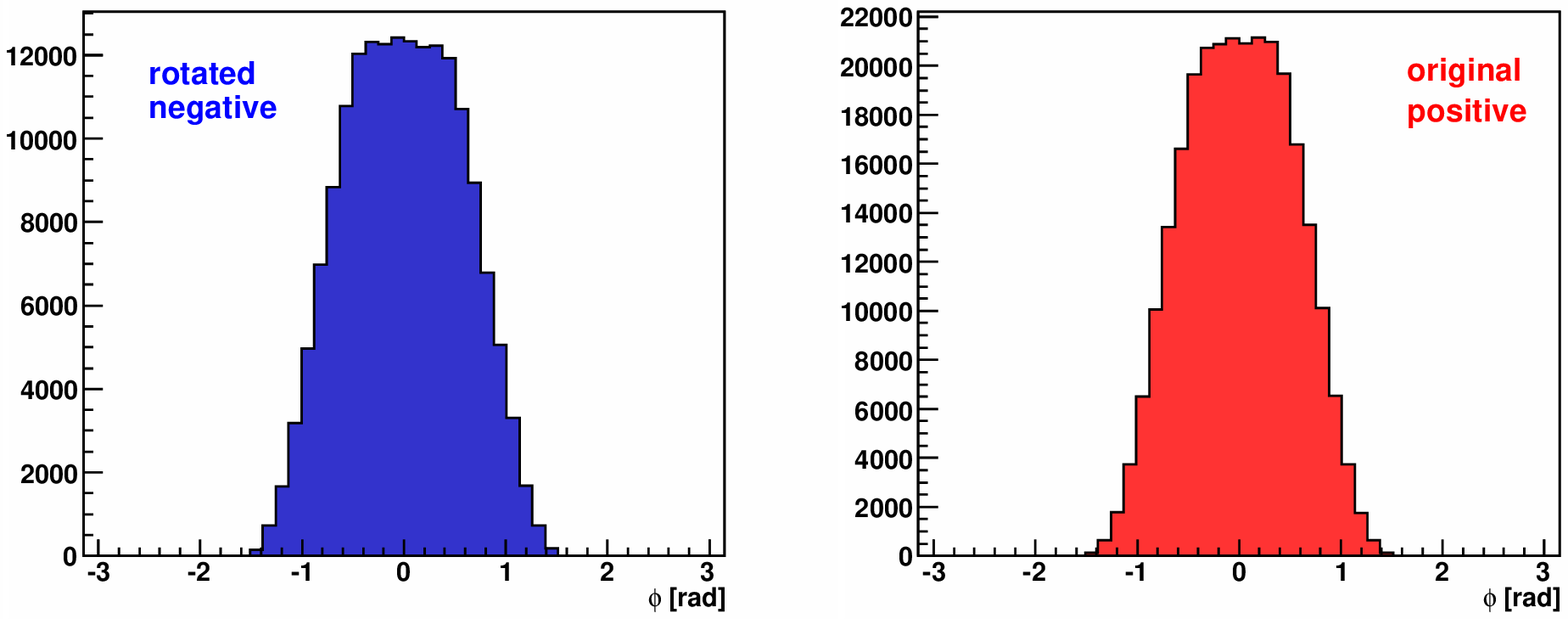}}
\vspace{-0.5pc}
\caption{\label{angle_shift}\small Examples of $\phi$ distributions at forward rapidity for positively and negatively charged particles before (upper plots) and after (lower plots) the shift.}
\end{minipage} 
\end{figure}

The NA49 detector has a limited azimuthal angle acceptance; the acceptance losses are concentrated mainly around the up-down regions. Due to the magnetic field in VTPC-1/2 particles of opposite charge are almost completely separated in the MTPCs (Fig. \ref{na49_az_acc}); the MTPCs cover mainly the forward rapidity region. 
As the detector is left-right symmetric, acceptance for positively and negatively charged particles is the same, provided the azimuthal angle for one charge is reflected. To allow a quantitative comparison of $\Phi_{\phi}$ for positively and negatively charged particles we rotate particles of one charge (Fig. \ref{angle_shift}).

\subsection{Preliminary results}

\begin{figure}[h]
\hspace{1pc}%
\begin{minipage}{17pc}
\centerline{\includegraphics[width=17pc]{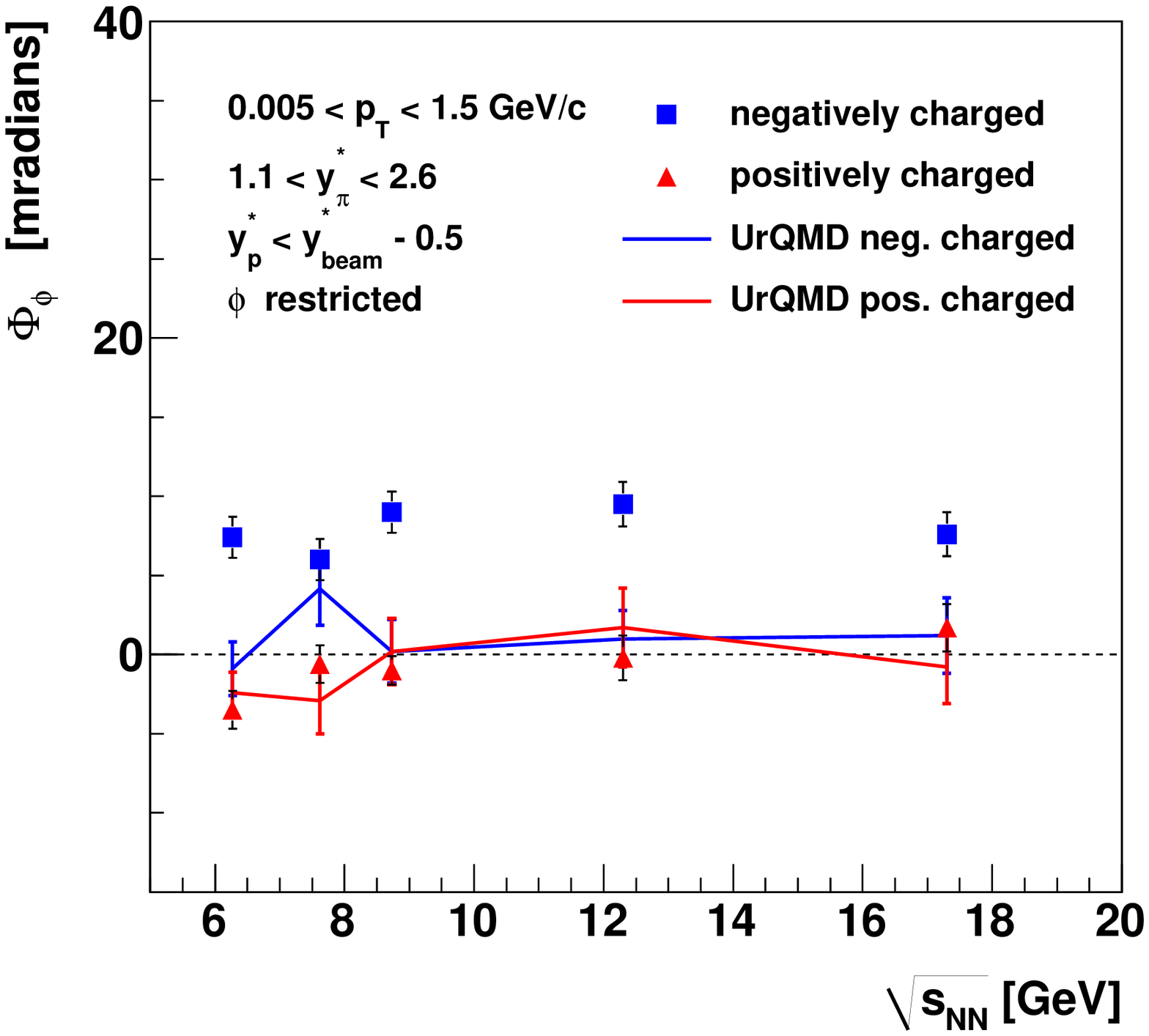}}
\vspace{-2pc}
\caption{\label{na49_energy}\small $\Phi_\phi$ as a function of energy. NA49 data compared to UrQMD 1.3 with the same kinematic cuts. See \cite{phipt_energy} for details of kinematic cuts and precise azimuthal acceptance description.}
\end{minipage}\hspace{2pc}%
\begin{minipage}{17pc}
\centerline{\includegraphics[width=17pc]{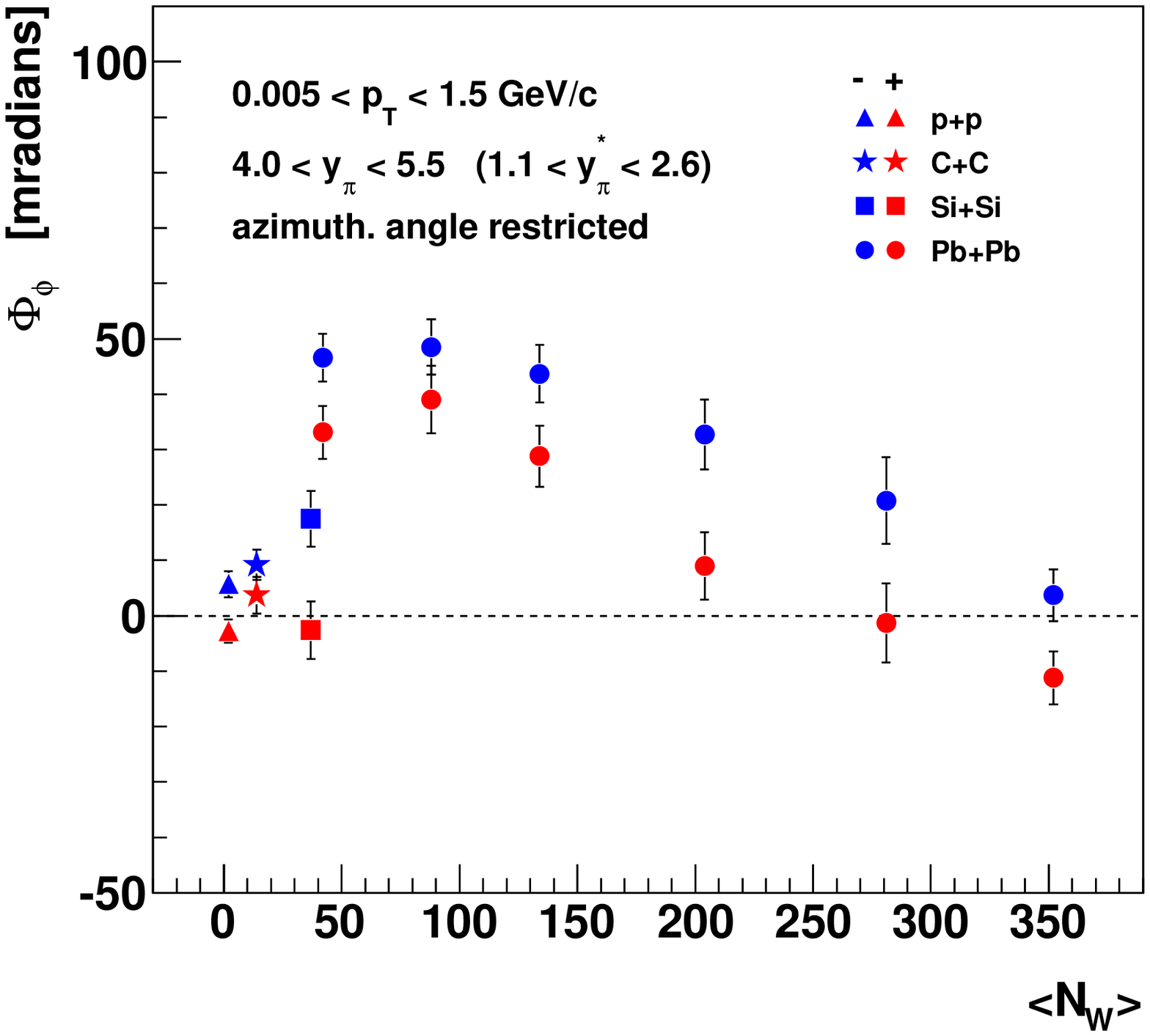}}
\vspace{-2pc}
\caption{\label{na49_size}\small $\Phi_\phi$ as a function of system size. See \cite{phipt_syssize} for details of kinematic cuts and precise azimuthal acceptance description. \newline \newline}
\end{minipage} 
\end{figure}

\begin{figure}[h]
\hspace{1pc}%
\begin{minipage}{17pc}
\centerline{\includegraphics[width=17pc]{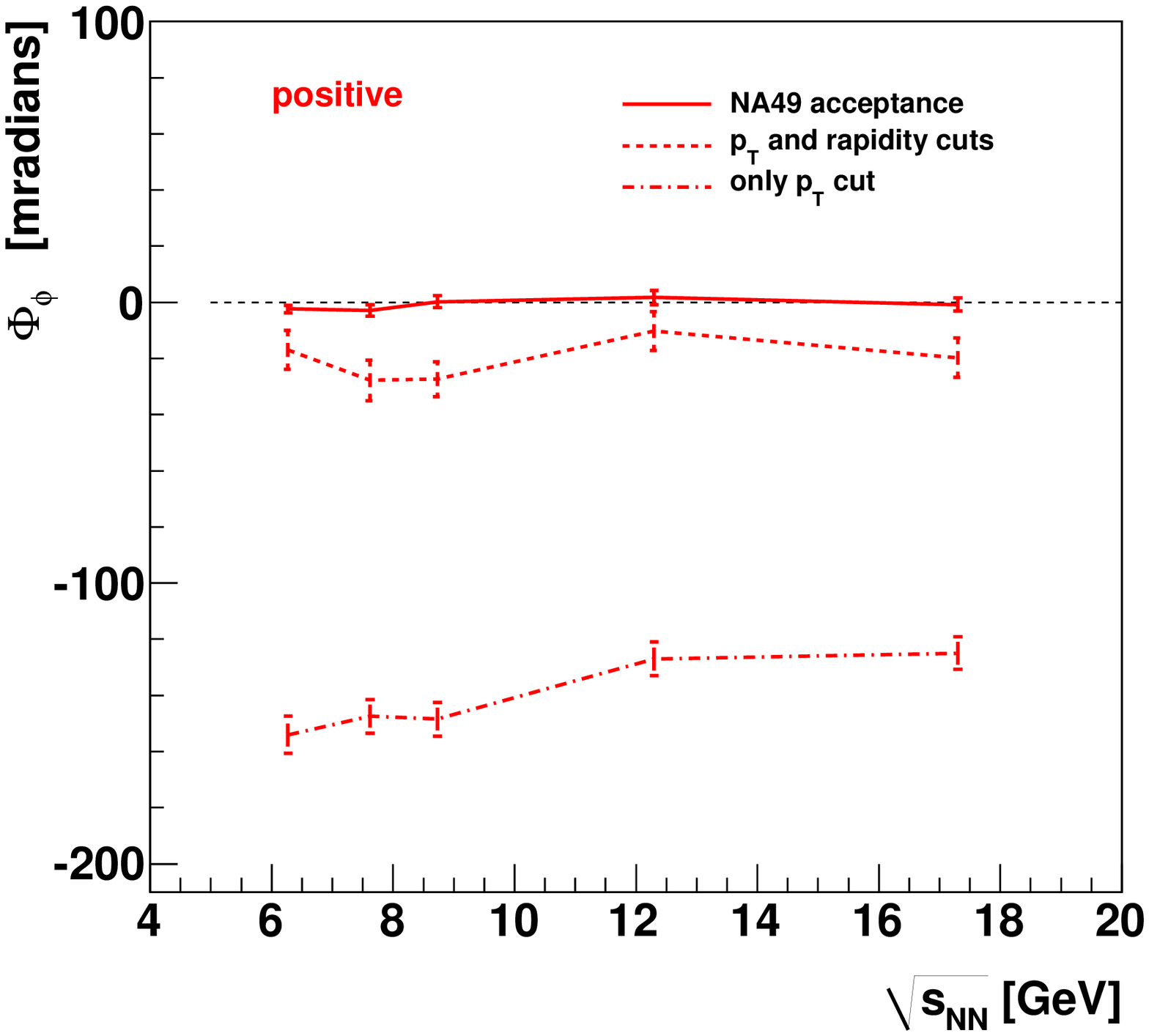}}
\vspace{-2pc}
\caption{\label{urqmd_13_pos}\small $\Phi_\phi$ (for pos. charged particles) as a function of energy for Pb+Pb collisions generated using UrQMD 1.3 with different kinematic restrictions.}
\end{minipage}\hspace{2pc}%
\begin{minipage}{17pc}
\centerline{\includegraphics[width=17pc]{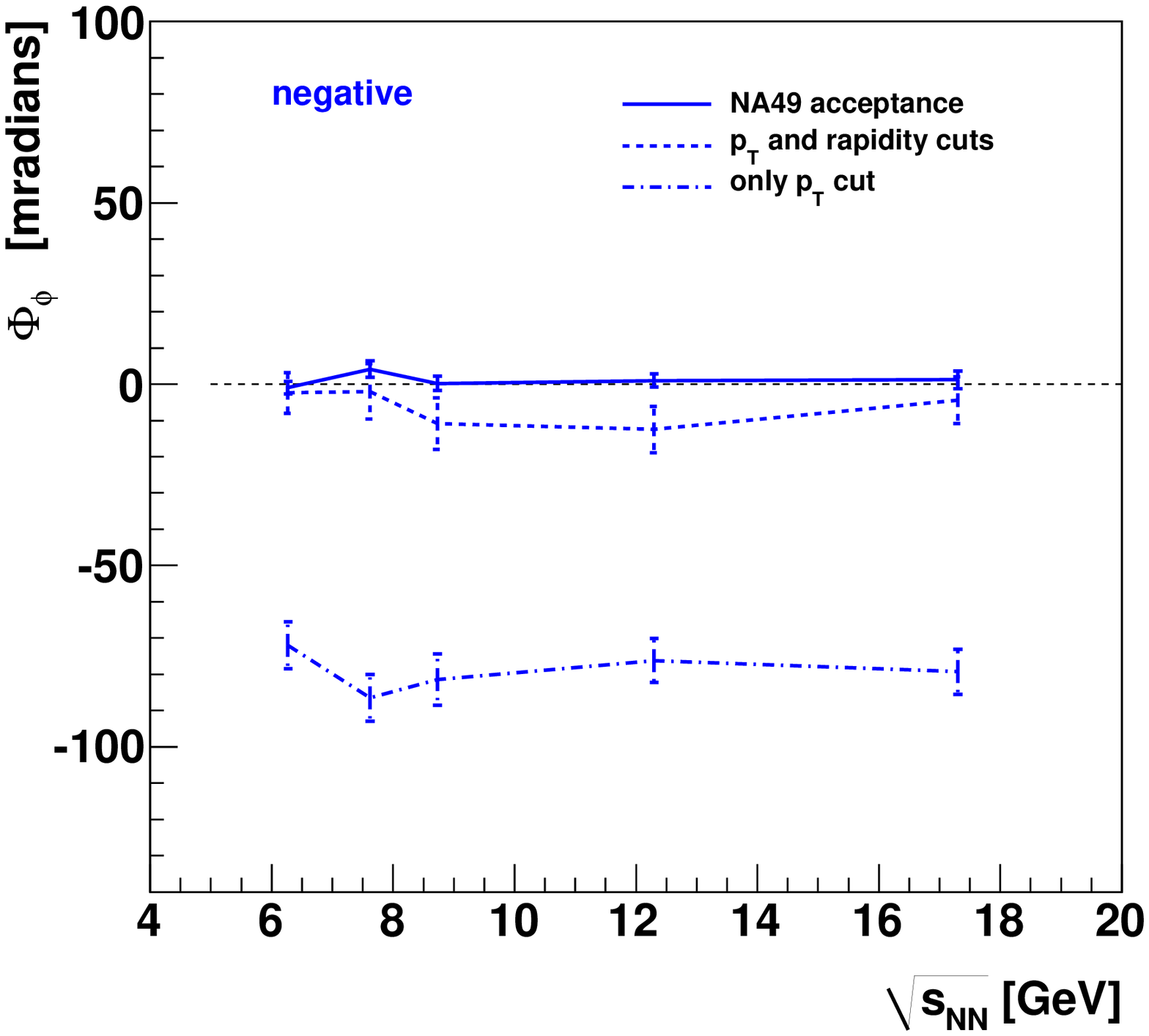}}
\vspace{-2pc}
\caption{\label{urqmd_13_neg}\small $\Phi_\phi$ (for neg. charged particles) as a function of energy for Pb+Pb collisions generated using UrQMD 1.3 with different kinematic restrictions.}
\end{minipage} 
\end{figure}

Figure \ref{na49_energy} shows the energy dependence of $\Phi_\phi$ for the 7.2\% most central Pb+Pb interactions. There is no significant energy dependence, however negatively charged particles show positive $\Phi_{\phi}$ values. The results from UrQMD, obtained with the same kinematic restrictions, show values consistent with zero. 
In the simulation the limited NA49 acceptance causes a major reduction in the
absolute values of the $\Phi_{\phi}$ measure (Figs. \ref{urqmd_13_pos} and \ref{urqmd_13_neg}; in Fig. \ref{urqmd_pp}, showing UrQMD results for p+p data, the complete rapidity and azimuthal angle range was used).
The main effect is due to the rapidity cut. The additional azimuthal angle restriction results in a $\Phi_{\phi}$ value consistent with zero.

Figure \ref{na49_size} presents the system size dependence of $\Phi_\phi$ at 158$A$ GeV (top SPS energy). Significant positive values of $\Phi_\phi$ are observed with a maximum for peripheral Pb+Pb interactions. This result is qualitatively similar to the results of multiplicity \cite{mryb} and average $p_T$ \cite{phipt_syssize} fluctuations in NA49. Further studies are planned to understand these intriguing effects.

\section{Summary}
Event-by-event azimuthal fluctuations were analyzed using the $\Phi_{\phi}$ measure, for models and for NA49 data. 
The properties of $\Phi_{\phi}$ were studied using toy models and UrQMD. Elliptic flow results in positive values of $\Phi_{\phi}$ which increase with the magnitude of flow and are sensitive to its fluctuations.
Contrary to flow, momentum conservation produces negative $\Phi_{\phi}$ values. The UrQMD model shows negative $\Phi_{\phi}$ values with a weak energy dependence.
The NA49 measurements for central Pb+Pb collisions showed weak energy dependence and positive $\Phi_{\phi}$ values for negatively charged particles. On the other hand, a significant system size dependence is observed at the highest SPS energy. This result is qualitatively similar to the behavior found for multiplicity \cite{mryb} and transverse momentum \cite{phipt_syssize} fluctuations.

\ack
{This work was supported by
the US Department of Energy Grant DE-FG03-97ER41020/A000,
the Bundesministerium fur Bildung und Forschung, Germany (06F~137),
the Virtual Institute VI-146 of Helmholtz Gemeinschaft, Germany,
the Polish Ministry of Science and Higher Education (1~P03B~006~30, 1~P03B~127~30, 
0297/B/H03/2007/33, N~N202~078735,  N~N202~078738, N~N202~204638),
the Hungarian Scientific Research Foundation (T032648, T032293, T043514),
the Hungarian National Science Foundation, OTKA, (F034707),
the Bulgarian National Science Fund (Ph-09/05),
the Croatian Ministry of Science, Education and Sport (Project 098-0982887-2878)
and Stichting FOM, the Netherlands.
}


\section*{References}
\begin{thebibliography}{9}
\bibitem{plasma_inst} Mr\'{o}wczy\'{n}ski S 1993 {\it Phys. Lett.} B {\bf 314} 118-21
\bibitem{Mrow_flow_fluct} Mr\'{o}wczy\'{n}ski S and Shuryak E 2003 {\it Acta Phys. Polon.} B {\bf 34} 4241-56
\bibitem{Mill_flow_fluct} Miller M, Snellings R arXiv:nucl-ex/0312008
\bibitem{Phi} Ga\'{z}dzicki M and Mr\'{o}wczy\'{n}ski S 1992 {\it Z. Phys.} C {\bf 54} 127
\bibitem{phipt_syssize} Anticic T et al. (NA49 Collab.) 2004 {\it Phys. Rev.} C {\bf 70} 034902
\bibitem{phipt_energy} Anticic T et al. (NA49 Collab.) 2009 {\it Phys. 
Rev.} C {\bf 79} 044904
\bibitem{delta_q} Alt C et al. (NA49 Collab.) 2004 {\it Phys. Rev.} C {\bf 70} 064903
\bibitem{mrow_phiphi} Mr\'{o}wczy\'{n}ski S {\it Acta Phys. Polon.} 2000 B {\bf 31} 2065
\bibitem{urqmd} Bass S A et al. 1998 {\it Prog. Part. Nucl. Phys.}
{\bf 41} 225-370;\\
Bleicher M et al. 1999 {\it J. Phys.} G {\bf 25} 1859-1896
\bibitem{pythia} Sjöstrand T, Mrenna S, Skands P arXiv:0710.3820
\bibitem{hijing} Wang X-N, Gyulassy M arXiv:nucl-th/9502021
\bibitem{na49_nim} Afanasiev S et al. (NA49 Collab.) 1999 {\it Nucl. 
Instrum. Meth.} A {\bf 430} 210
\bibitem{mryb} Alt C et al. (NA49 Collab.) 2007 {\it Phys. Rev.} C
{\bf 75} 064904

\end{thebibliography}
\end{document}